\documentclass[prd,showpacs,preprintnumbers,nofootinbib,eqsecnum]{revtex4}

\usepackage[dvipdfm]{graphicx}
\usepackage{dcolumn}
\usepackage{bm}
\usepackage{latexsym}
\usepackage{amsfonts}
\usepackage{amssymb}
\usepackage{amsmath}
\usepackage[usenames]{color}

\begin{document}

\preprint{KUNS-2369}

\title{
Exact Coleman-De Luccia Instantons
}
\author{Sugumi Kanno$^{1}$}
\author{Jiro Soda$^{2}$}
\affiliation{$^{1}$Institute of Cosmology, Department of Physics and Astronomy, 
Tufts University, Medford, Massachusetts 02155, USA \\
 $^{2}$Department of Physics,  Kyoto University, Kyoto 606-8502, Japan\\
}%

\date{\today}

\begin{abstract}
We present exact Coleman-De Luccia (CDL) instantons, which describe vacuum decay from Anti de Sitter (AdS) space, de Sitter (dS) space and Minkowski space to AdS space. We systematically obtain these exact solutions by considering deformation of Hawking-Moss (HM) instantons. We analytically calculate the action of instantons and discuss a subtlety in calculation of decay rates.
\end{abstract}

\maketitle
\section{Introduction}

 The development of string theory has revealed a vacuum structure of the universe, so-called ``the string theory landscape" where many metastable vacuum 
states exist. Then, the study of CDL instantons~\cite{Coleman:1980aw} describing the decay of such metastable states has attracted much interest recently. 
The vacuum decay in a fixed background has been well understood thanks to the seminal papers by Coleman and his collaborators~\cite{Coleman:1977py,Callan:1977pt}. 
 However, tunneling in the presence of gravity is still controversial. For example, the existence of negative modes may be gauge dependent for CDL instanton solutions~\cite{Tanaka:1992zw,Tanaka:1998mp,Tanaka:1999pj,Khvedelidze:2000cp}. Hence, interpretation of the CDL instantons is not so clear~\footnote{ 
 In the case of dS vacuum decay, there is an argument based on the WKB approximation to support the prescription using the CDL instantons~\cite{Gen:1999gi}.}.
 Moreover, there exist some claims on the string theory landscape that contradict each other. It is claimed that the eternal inflation may populate the string theory landscape~\cite{Brown:2011ry}. On the other hand, it is insisted that one of the effects of gravitational backreaction might be missing in the thin-wall approximation for the CDL instantons~\cite{Copsey:2011zj}. Then the decay rate is enhanced if we take into account the effect. It implies that most of vacua rapidly decay  and the string theory landscape would not be relevant to cosmology. Furthermore, there is a dispute on the 
 interpretation of the decay rate of the metastable vacuum into an AdS vacuum~\cite{Dvali:2011wk,Garriga:2011we}.
  Thus, it is worth pursuing the study of the vacuum decay with gravity.

It would be beneficial if we had an analytic treatment of the CDL instanton solutions describing the vacuum decay with gravity. So far, the only available tool has been the
 thin-wall approximation. However, since the string theory landscape has a variety of vacua, there is a limitation of the thin-wall approximation. Of course, one can resort to the numerical analysis. However, it is difficult to grasp the physics behind numerical calculations. Therefore, we need analytical tools beyond the thin-wall approximation. In fixed background cases, it is not so difficult to construct
exactly solvable models~\cite{Hamazaki:1995dy,Koyama:1999ai,Pastras:2011zr,Dutta:2011ej,Dutta:2011rc}.
Recently, an attempt to construct analytic CDL solutions has been performed~\cite{Dong:2011gx}. The method is well known in the literature of braneworld
~\cite{DeWolfe:1999cp,Gremm:1999pj,Csaki:2000fc,Kobayashi:2001jd}.
However, they did not get analytic forms of potential functions~\cite{Dong:2011gx}. In this paper, we present exact CDL instantons for the first time by focusing
 on the HM instantons~\cite{Hawking:1981fz}. The HM instantons, which are nothing but Euclidean dS solutions, 
play an important role when the CDL instantons disappear. 
When the curvature scale of the potential barrier is large compared to the scale determined by the potential energy, CDL instantons exist. As the curvature scale becomes small, 
however, the CDL instantons approach and eventually merge into the HM instantons.
Hence, it is expected that we can construct the CDL instantons by making deformation of
the Euclidean dS solutions. Indeed, we find many exact CDL instantons using this strategy.
 In particular, we find potentials for which we can construct exact CDL instantons. Using our exactly solvable models, we can calculate decay rates analytically.   

So far, vacuum decays from dS to dS and from dS to Minkowski have been well discussed in conjunction with cosmology. However, recent string theory landscape picture tells us the importance of AdS vacua.  Hence, our primary interest in this paper is the vacuum decay rate to the AdS vacuum.  We note that Banks argued that the decay to an AdS crunch is an actual fate of the decay process~\cite{Banks:2002nm}. Using our exactly solvable models, we investigate various decay processes. In addition to these practical calculations, we will discuss a subtle point related to the interpretation of CDL instantons.

The organization of this paper is as follows. In section II, we give basic equations and a strategy to find exact solutions.  In section III, we present exact solutions which describe vacuum decay from AdS space, dS space and Minkowski space to AdS space. In section IV, we calculate the action of instantons. In section V, we evaluate decay rates for various decay processes. 
We find there exists a subtlety in evaluating decay rates simply by following a CDL prescription. 
 Section VI is devoted to conclusion.

\section{Coleman-De Luccia Set up}

Suppose that there are two perturbatively stable vacua and one of them is a 
metastable false vacuum. The false vacuum will decay into a true vacuum.
The decay process can be described by an instanton in a semi-classical 
approximation. In the absence of gravity, a method for the instanton 
is well established. On the other hand, in the presence of gravity, 
it is believed that CDL instantons play an important role. 
In this section, we set up basic equations for obtaining the CDL
instantons and describe a strategy to get exact solutions.

We begin with $d$-dimensional Euclidean action for the 
gravitational field $g_{\mu\nu}$ and the scalar field $\phi$:
\begin{eqnarray}
S_{E} =-\frac{1}{2\kappa^2}\int_{M} d^dx\sqrt{g}~R
- \frac{1}{\kappa^2} \int_{\partial M} d^{d-1}x \sqrt{h}~K
+\int_{M} d^dx \sqrt{g}\left[~
\frac{1}{2}g^{\mu\nu}\partial_\mu\phi\partial_\nu\phi
+V(\phi)
~\right]\,,
\end{eqnarray}
where $\kappa^2=8\pi G$, $R$ is the Ricci scalar for the metric $g_{\mu\nu}$, 
$K$ is the trace part of the extrinsic curvature and $h$ is the determinant of
an induced metric on the boundary.
The second term is the Gibbons-Hawking (GH) term which is necessary to make the
variational principle consistent~\cite{Gibbons:1976ue}. 
This action is well defined for spatially compact geometries but diverges
for non-compact ones.
Hence, the physical action is required to take the form~\cite{Hawking:1995fd} 
\begin{eqnarray}
I = S_{E} \left[~ g_{\mu\nu} ,~\phi ~\right] 
          - S_b \left[~ g_{b \mu\nu} ,~\phi_b ~\right] \ ,
          \label{physical}
\end{eqnarray}
so that the physical action of the reference background becomes zero. 
Here, suffix $b$ represents the background. 

According to this prescription, the appropriate action for asymptotically flat space is
expressed by
\begin{eqnarray}
I =-\frac{1}{2\kappa^2}\int_{M} d^dx\sqrt{g}~R
- \frac{1}{\kappa^2} \int_{\partial M} d^{d-1}x \sqrt{h}~[K - K_0]
+\int_{M} d^dx \sqrt{g}\left[~
\frac{1}{2}g^{\mu\nu}\partial_\mu\phi\partial_\nu\phi
+V(\phi)
~\right] \ ,
\label{physical2}
\end{eqnarray}
where $K_0$ is the trace of the extrinsic curvature of the boundary
embedded in flat
 space~\cite{Gibbons:1976ue,Hawking:1995fd}.
Apparently, this action vanishes for the flat space.
For other backgrounds, we need to go back to the more general form in 
Eq.~(\ref{physical}).
In the process of a vacuum decay, 
the choice of $S_b$ is a subtle issue as we will see later in Section 
\ref{subtlety}.

It is believed that the vacuum decay is described by $O(d)$ symmetric 
instantons. Then, we take the metric with $O(d)$ symmetry of the form
\begin{eqnarray}
ds^2={\cal N}(\xi)^2d\xi^2 + a(\xi)^2d\Omega^2_{d-1}  \ . 
\end{eqnarray}
Here, ${\cal N(\xi)}$ and $a(\xi)$ are the lapse function and the scale
factor of the coordinate $\xi$, respectively. 
$d\Omega^2_{d-1}$ is the line element of a unit $(d-1)$-dimensional sphere. 
The action with this metric is given by
\begin{eqnarray}
S_{E} =-\frac{(d-1)(d-2)}{2\kappa^2} v_{S^{d-1}}\int d\xi
\left[~
\frac{1}{\cal N}a^{d-3}a^{\prime 2}+{\cal N}a^{d-3}
~\right]
+ v_{S^{d-1}}\int d\xi~a^{d-1}\left[~
\frac{1}{2{\cal N}}\phi^{\prime 2} + {\cal N}V
~\right]   \ ,
\label{action}
\end{eqnarray}
where a prime denotes derivative with respect to $\xi$ and 
$v_{S^{d-1}}$ is a volume of the unit $(d-1)$-dimensional
sphere. Note that the internal scalar curvature of a unit sphere is 
$R^{(d-1)}=(d-1)(d-2)$. The GH term canceled the surface
term of the Einstein-Hilbert action.

Varying the action with respect to ${\cal N}$ and ${\phi}$ gives
the following basic equations. We set ${\cal N}=1$ after the variation. 
The Hamiltonian constraint equation is given by
\begin{eqnarray}
\left(\frac{a^\prime}{a}\right)^2=\frac{1}{a^2}
+\frac{2\kappa^2}{(d-1)(d-2)}\left(
\frac{1}{2}\phi^{\prime 2}-V
\right)   \ .
\label{hc1}
\end{eqnarray}
The scalar field equation becomes
\begin{eqnarray}
\phi^{\prime\prime}+(d-1)\frac{a^\prime}{a}\phi^\prime
-\frac{dV}{d\phi}=0  \ .
\label{sc1}
\end{eqnarray}
The equation obtained by taking the variation with respect to $a(\xi)$ is 
redundant.
By taking the derivative with respect to $\xi$ of Eq.~(\ref{hc1})
and plugging the Eq.~(\ref{sc1}) into it, we find the derivative
of the scalar field is written only by the scale factor. Furthermore, putting
the result back into Eq.~(\ref{hc1}), we get the potential for the scalar
field expressed only by the scale factor as well. Then the basic equations become
\begin{eqnarray}
\frac{\kappa^2}{d-2}\phi^{\prime 2}&=&
\left(\frac{a^\prime}{a}\right)^2-\frac{1}{a^2}
-\frac{a^{\prime\prime}}{a}  \ ,
\label{hc2}\\
\frac{2\kappa^2}{(d-2)^2}V(\phi)&=&
\frac{1}{a^2}-\left(\frac{a^\prime}{a}\right)^2
-\frac{1}{d-2}\frac{a^{\prime\prime}}{a}   \ .
\label{sc2}
\end{eqnarray}
Thus, if we put a desired function for $a(\xi)$ in these equations,
both of the derivative of the scalar field and the potential of the scalar 
field are given as a function of $\xi$ from Eqs.~(\ref{hc2}) and (\ref{sc2}).
Combining those results, we then get the form of $V(\phi)$. The potential we
look for is to have metastable and stable vacuum points and the metastable 
vacuum 
will eventually decay into the true vacuum.

In order to find CDL instanton solutions, it is convenient to move to the conformal coordinate defined by $d\xi=a(z)dz$, which corresponds to the choice ${\cal N}=a$. Then, Eqs.~(\ref{hc2}) and (\ref{sc2}) become
\begin{eqnarray}
\frac{\kappa^2}{d-2}\left(\frac{d\phi}{dz}\right)^2
&=&\frac{2}{a^2}\left(\frac{da}{dz}\right)^2
-\frac{1}{a}\frac{d^2a}{dz^2}-1  \ ,
\label{hc3}\\
\frac{2\kappa^2}{(d-2)^2}V(\phi)&=&
\frac{1}{a^2}\left[~
1-\frac{d-3}{d-2}\frac{1}{a^2}\left(\frac{da}{dz}\right)^2
-\frac{1}{d-2}\frac{1}{a}\frac{d^2a}{dz^2}
~\right]  \ . 
\label{sc3}
\end{eqnarray}

In the next section, we will present a deformation method for obtaining exact 
CDL instantons using this strategy.
It is not difficult to perform our analysis in arbitrary dimensions, however,
 we will concentrate on exact solutions in four dimensions $(d=4)$ 
for simplicity.

\section{Exactly Solvable Models }

In this section, we present exactly solvable models of the vacuum decay with gravity.
 As we explained in the previous section, we give some scale factors 
$a(z)$ first. As a key to find exact solutions, we focus on the
HM instantons which are Euclidean  dS 
solutions~\cite{Hawking:1981fz}. 
If the potential satisfies the condition $3 |V''| > 4 \kappa^2 V$ at the maximal
point of the potential barrier, 
both of the CDL and HM instantons exist~\cite{Jensen:1983ac}. In this case, 
the CDL instantons describe the decay of a false vacuum. However, if the
potential does not satisfy this condition, the CDL instantons do not exist 
anymore \footnote{Strictly speaking, this is not always true. The existence
of the CDL instantons depends on the higher order derivatives of the potential
~\cite{Hackworth:2004xb}.}. Instead, the HM instantons describe the decay of the false vacuum. 
This implies that the CDL instantons 
approach to the HM instantons as $3|V''|$ approaches $4\kappa^2 V$, and 
eventually degenerate into the HM instantons. There the scale factor is
given by $ a(z)= \ell_{\rm dS} /\cosh z$, where $\ell_{\rm dS}$ is the
curvature radius in the Euclidean dS spaces. Since the HM instanton is 
a limit of the CDL instanton, we can make use of deformation of HM 
instantons for obtaining the CDL instantons. Namely, we can put the ansatz
 for the scale factor 
\begin{eqnarray}
a(z) = \ell~\frac{f\left(\tanh z\right)}{\cosh z} \ ,
\label{scalefactor}
\end{eqnarray}
where we refer a function of $\tanh z$, $f$, as the deformation function 
from the HM instantons. Here, $\ell$ is a length scale in the system. 
This function guarantees a compact asymptotic 
structure at $z\rightarrow\infty$.
Substituting this into Eq.~(\ref{hc3}), we obtain 
\begin{eqnarray}
\frac{\kappa^2}{4} \left(\frac{d\phi}{d\lambda}\right)^2
= \left(\frac{1}{f}\frac{df}{d\lambda}\right)^2 -\frac{1}{2f}\frac{d^2 f}{d\lambda^2}\,,
\label{hc4}
\end{eqnarray}
where we defined a variable $\lambda = \tanh z$. 
Using the function $f$, we can write the potential as
\begin{eqnarray}
\frac{\kappa^2\ell^2}{2} V
= \frac{1}{f^2}\left[~ 
\frac{3}{2}-\frac{1}{2}\left(1-\lambda^2\right)
\frac{1}{f}\frac{d^2 f}{d\lambda^2} 
    + 3\lambda~\frac{1}{f} \frac{df}{d\lambda} 
    -\frac{1}{2}\left(1-\lambda^2\right) 
\left( \frac{1}{f}\frac{df}{d\lambda}\right)^2 
~\right]  \ .
\label{sc4}
\end{eqnarray}
Now, we can discuss various exact solutions using the above formulas.

\subsection{ From Runaway State to AdS}
\label{IIIA}

\subsubsection{Type $f(\lambda ) = \exp \left( \alpha \lambda \right)$}

\begin{figure}[ht]
\begin{center}
\begin{minipage}{8.5cm}
\begin{center}
\hspace{-1.5cm}
\includegraphics[width=95mm]{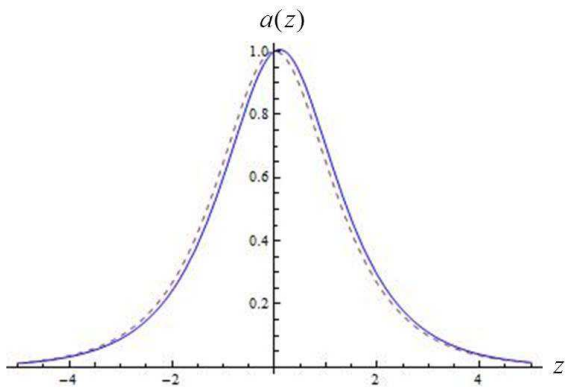}\vspace{0cm}
\caption{The scale factor as a function of $z$ with $\ell=1$. 
A solid line
corresponds to $\alpha=0.1$. 
A dashed line corresponds to the dS solution $\alpha=0$.}
\label{fig:1}
\end{center}
\end{minipage}
\hspace{1mm}
\begin{minipage}{8.5cm}
\begin{center}
\vspace{-4mm}
\hspace{-1.5cm}
\includegraphics[width=95mm]{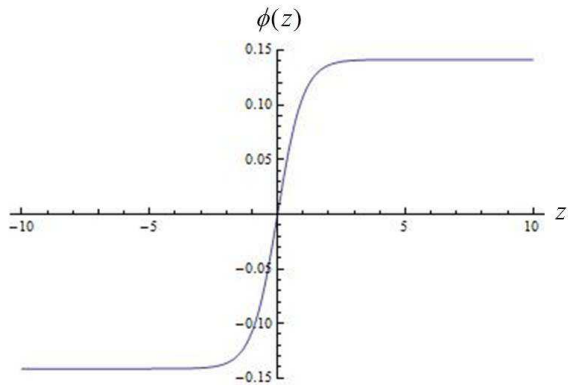}\vspace{0cm}
\caption{The scalar field as a function of $z$ in the case of 
$\alpha=0.1$. $\kappa=1$.}
\label{fig:2}
\end{center}
\end{minipage}
\end{center}
\end{figure}

Looking at Eq.~(\ref{hc4}), we find 
a kink type solution proportional to $\lambda$.
 Indeed,  we can choose the deformation function as
\begin{eqnarray}
f(\lambda ) = \exp \left( \alpha \lambda \right) \ ,
\label{f3}
\end{eqnarray}
which leads to the scale factor
\begin{eqnarray}
 a (z) = \ell~\frac{e^{\alpha \tanh z}}{\cosh z} \ ,
\label{a3}
\end{eqnarray}
where $\alpha$ is a constant parameter. This scale factor
 has a profile deformed from Euclidean dS spaces, $\alpha =0$, 
(see Fig.~\ref{fig:1} ).
Interestingly, it turns out that this
simple deformation changes the constant potential corresponding to the Euclidean dS space into a runaway potential. Plugging Eq.~(\ref{f3}) into Eq.~(\ref{hc4}) and 
integrating it, 
we get the scalar field 
\begin{eqnarray}
\frac{\kappa}{2}\phi = \frac{\alpha}{\sqrt{2}} \lambda
                    = \frac{\alpha}{\sqrt{2}} \tanh z \ ,
\label{phi-profile-1}
\end{eqnarray}
where we took a plus sign of $\phi$ and put the constant of integration zero. 
A minus sign corresponds to a parity flip $z\rightarrow -z$. 
We have plotted the profile of the scalar field in Fig.~\ref{fig:2}.
Since $|\tanh z|<1$, the range of the scalar field is restricted to
$-\frac{\sqrt{2}\alpha}{\kappa}<\phi<\frac{\sqrt{2}\alpha}{\kappa}$.
The CDL instanton is restricted to this range in this way.

\begin{figure}[htbp]
\includegraphics[width=10cm]{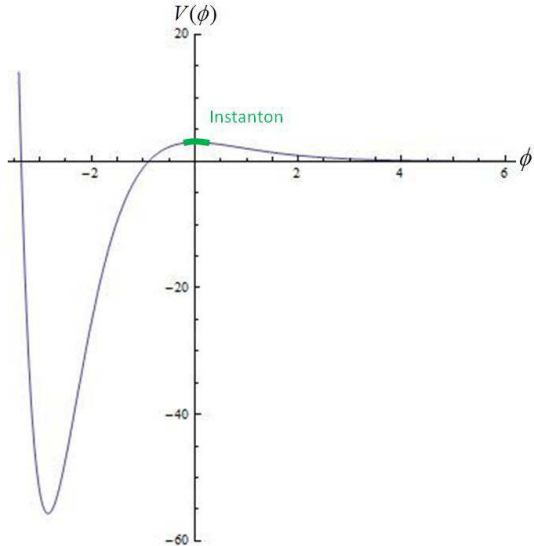}
\caption{The potential as a function of $\phi$ in the case of $\alpha=0.1$. 
$\kappa=\ell=1$.
The true vacuum is an AdS vacuum.
The metastable false vacuum is a Minkowski vacuum at the infinity of $\phi$. 
The instanton describing the vacuum decay is shown in a green line.}
\label{fig:3}
\end{figure}

Putting the scale factor Eq.~(\ref{a3}) into Eq.~(\ref{sc4}), 
we can calculate the potential $V$ as a function of $z$
\begin{eqnarray}
\kappa^2\ell^2~V = \frac{1}{2\cosh^2 z}e^{-2\alpha \tanh z} 
\left(~3-4\alpha^2 + 3\cosh 2z + 6\alpha \sinh 2z~\right) \ .
\label{V-exp}
\end{eqnarray}
As we can solve $\tanh z$ as a function of $\phi$ from Eq.~(\ref{phi-profile-1}), 
if we substitute it into Eq.~(\ref{V-exp}), the potential $V$
is obtained as a function of $\phi$
\begin{eqnarray}
\kappa^2\ell^2~V(\phi) = e^{-\sqrt{2}\kappa\phi}
\left(~3-2\alpha^2 + 3\sqrt{2}\kappa\phi + \kappa^2\phi^2 ~\right) \ .
\end{eqnarray}
The potential has a runaway shape as in Fig.~\ref{fig:3}, namely, there is no
true minimum in the false vacuum side. In Fig.~\ref{fig:3}, 
a green line shows the range of the CDL instanton. This instanton solution can be interpreted as a decay process from the runaway 
state to an AdS vacuum. 
After the tunneling, the scalar field will roll down the potential and follows Lorentzian equations. 

Note that we can deform the potential keeping the shape of the green line
fixed. This means that we can also interpret that the instanton mediates the
tunneling between two dS vacua in the case of Fig.~\ref{fig:3}.

\subsubsection{Type $ f (\lambda) = 1+\beta\lambda$}

\begin{figure}[ht]
\begin{center}
\begin{minipage}{8.5cm}
\begin{center}
\hspace{-1.5cm}
\includegraphics[width=10cm]{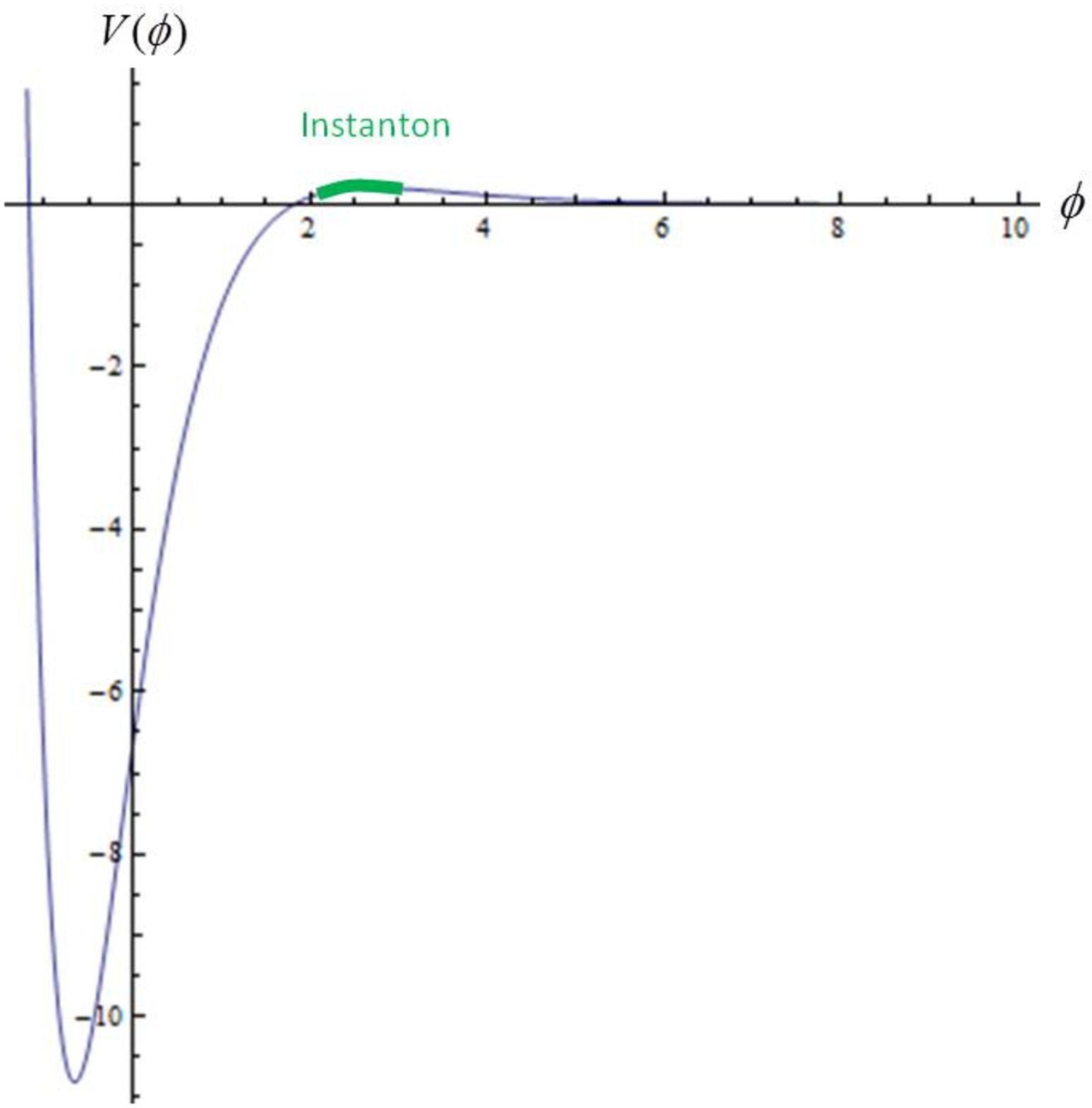}
\caption{The potential, $V$  versus $\phi$ 
in the case of $\ell=3.8,~\ell\beta=0.8$ in type 
$ f (\lambda) = 1+\beta\lambda$ model. $\kappa=1$.
A green line shows by the CDL instanton. }
\label{fig:4}
\end{center}
\end{minipage}
\hspace{1mm}
\begin{minipage}{8.5cm}
\begin{center}
\hspace{-1.5cm}
\includegraphics[width=10cm]{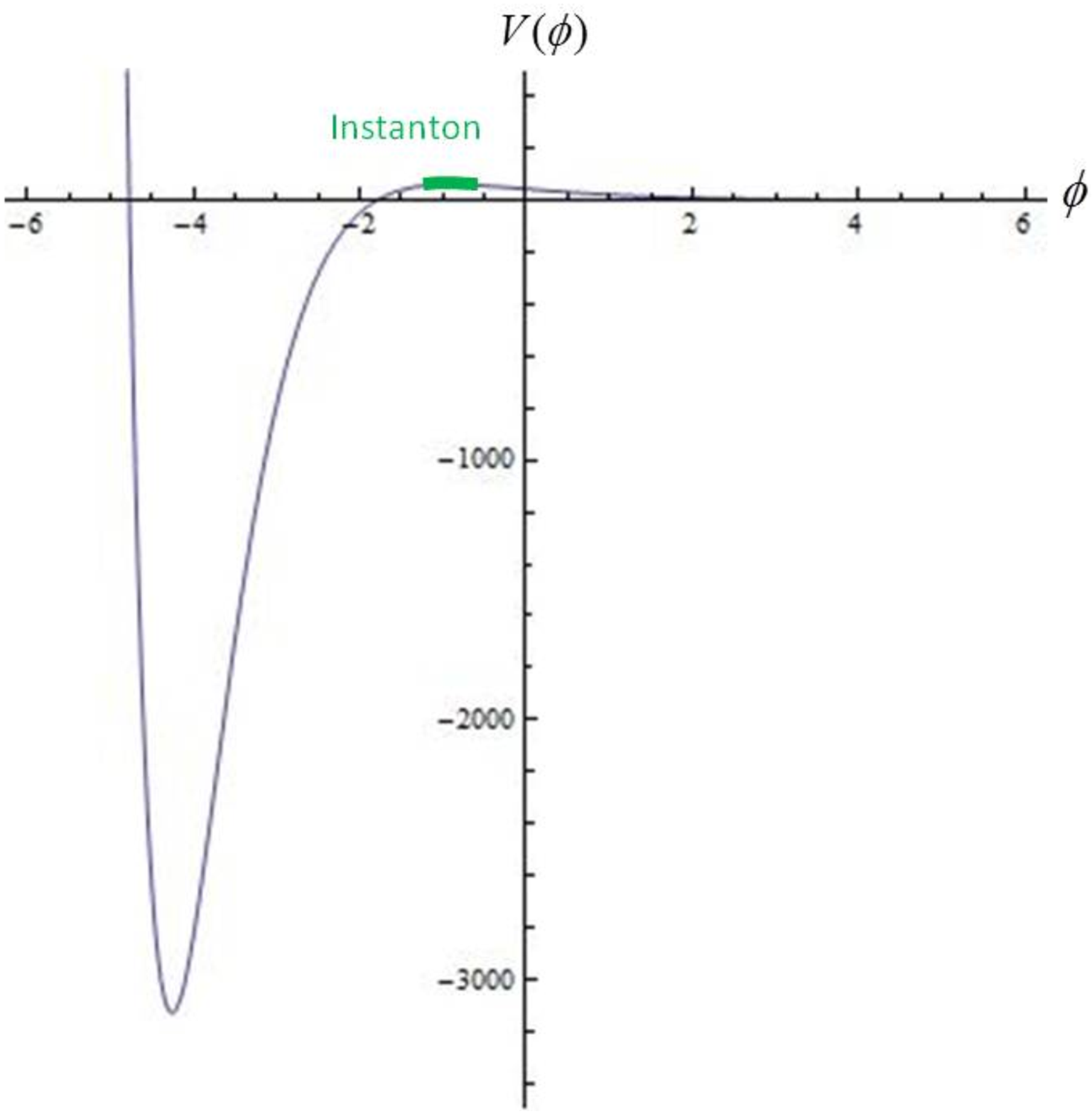}
\caption{The potential, $V$  versus $\phi$ in the case of 
$\alpha=3,~\beta=1,~\ell=2,~\ell\gamma=1$ in type 
$f(\lambda) = \left(1 + \gamma\lambda\right)
/\left(\alpha + \beta \lambda\right)^2$ model. $\kappa=1$.
The green line shows the CDL instanton.}
\label{fig:5}
\end{center}
\end{minipage}
\end{center}
\end{figure}

Another simple choice is to take $d^2 f /d\lambda^2 =0$ in Eq.~(\ref{hc4}). 
Then the deformation function becomes
\begin{eqnarray}
 f (\lambda) = 1+\beta\lambda \ ,
 \label{f1}
\end{eqnarray} 
where $\beta $ is a constant parameter. 
The scale factor is then given by 
\begin{eqnarray}
a(z)=\ell~\frac{1+\beta\tanh z}{\cosh z} \ .
\label{a1}
\end{eqnarray}
Here, we have to impose the condition $0 <\beta <1$ to get a positive scale factor $a$.
Changing the sign of $\beta$ corresponds to a parity flip $z \rightarrow -z$.  
Putting Eq.~(\ref{f1}) into Eq.~(\ref{hc4}) and integrating it over, we obtain
the scalar field of this form
\begin{eqnarray}
\frac{\kappa}{2}\phi = \log \left( 1 + \beta \lambda \right) 
= \log \left( 1 + \beta \tanh z \right) \ .
\label{phi1}
\end{eqnarray}
Since $|\tanh z|<1$, the range of the scalar field is
restricted to $1-\beta < e^{\frac{\kappa}{2}\phi}<1+\beta$.
Using Eq.~(\ref{sc4}), 
we can also calculate the potential $V$ as a function of $z$
\begin{eqnarray}
\kappa^2\ell^2~V
= \frac{3}{\left(1+\beta\tanh z\right)^2}
-\frac{\beta^2\left(1-\tanh^2 z\right)}{\left(1+\beta\tanh z\right)^4}
+\frac{6\beta\tanh z}{\left(1+\beta\tanh z\right)^3} \ .
\label{V1}
\end{eqnarray}
As we can solve $\tanh z$ as a function of $\phi$ from Eq.~(\ref{phi1}), 
if we substitute it into Eq.~(\ref{V1}), we can deduce the potential $V$
as a function of $\phi$
\begin{eqnarray}
\kappa^2\ell^2~V(\phi)
=  10e^{-\kappa\phi}-8 e^{-\frac{3}{2}\kappa\phi}
+\left(1-\beta^2\right)e^{-2\kappa\phi}   \ .
\end{eqnarray}
With the parameters $\ell =3.8 \ , \ell\beta =0.8$, 
we have the similar potential
as the previous case as in Fig.~\ref{fig:4}. 
The green line shows the range of the instanton. 
We note that this type of potential often appears 
in flux compactification models. 
In the context of the flux compactification, 
the scalar field $\phi$ plays a roll of a radion field and
runs away towards the Minkowski vacuum $\phi \rightarrow \infty$,
which is interpreted as a de-compactification in flux compactification models~
\cite{BlancoPillado:2009di,BlancoPillado:2009mi}.

\subsubsection{Type 
$f(\lambda) = \left(1 + \gamma \lambda\right)
/\left(\alpha + \beta \lambda\right)^2$}
\label{type3}

As an extension of the form Eq.~(\ref{f1}), we find 
\begin{eqnarray}
f(\lambda) = \frac{1 + \gamma \lambda}{\left(\alpha + \beta\lambda\right)^2}
\end{eqnarray}
is also solvable. Here, $\alpha , \beta $ and $\gamma$ are constant 
parameters. 
The scale factor reads
\begin{eqnarray}
a(z)=\frac{1+\gamma\tanh z}{\left(~\alpha+\beta\tanh z~\right)^2}
\cdot\frac{\ell}{\cosh z} \ .
\label{aa3}
\end{eqnarray}
Here, we have to impose the condition $0< \gamma <1$ and $|\beta|<\alpha$ 
to get a positive scale factor $a$.
 Eq.~(\ref{hc4}) gives a simple expression
\begin{eqnarray}
 \frac{\kappa}{2} \frac{d\phi}{d\lambda} 
= \frac{\alpha \gamma -\beta}{\left(\alpha + \beta \lambda\right)\left(1 + \gamma \lambda\right)} \ .
\end{eqnarray}
Note that we took one of sign of $\pm$ of this result simply because
the other sign just changes $\phi$ to $-\phi$. 
We can integrate it such as
\begin{eqnarray}
\frac{\kappa}{2}\phi=\log\left(
\frac{1+\gamma\tanh z}{\alpha+\beta\tanh z}
\right) \ .
\label{phi2}
\end{eqnarray}
Since $|\tanh z|<1$, the region of the scalar field is
restricted to $
\frac{2}{\kappa}\log\big|
\frac{1-\gamma}{\alpha-\beta}\big|
<\phi<\frac{2}{\kappa}\log\big|
\frac{1+\gamma}{\alpha+\beta}\big|$ (see Fig.~\ref{fig:5} ).
Thus the potential is obtained in the same way as previous cases
\begin{eqnarray}
\kappa^2\ell^2 V(\phi) =
10\left(\alpha^2-\beta^2\right)e^{-\kappa\phi}
-8\left(\alpha-\beta\gamma\right)e^{-\frac{3}{2}\kappa\phi}
+\left(1-\gamma^2\right)e^{-2\kappa\phi}  \ ,
\end{eqnarray}
where we used the inversion $\tanh z=\frac{1-\alpha e^{\frac{\kappa}{2}\phi}}
{\beta e^{\frac{\kappa}{2}\phi}-\gamma}$ in Eq.~(\ref{phi2}).
We get a runaway type potential with an AdS minimum again. 

We see that those instantons do not tunnel from a local minimum to a 
local minimum of the potential. This is the effects of gravity. Thus it is difficult to read off 
the false and true vacuum of the potential without knowing the whole shape of it. 

\subsection{From vacuum to vacuum}

Let us obtain the CDL instantons for the potentials having 
two  vacua. For this purpose, it is useful to rewrite the
right hand side of Eq.~(\ref{hc4}) as
\begin{eqnarray}
\frac{\kappa^2}{4} 
\left(\frac{d\phi}{d\lambda}\right)^2
= \frac{f}{2} \frac{d^2}{d\lambda^2} \left(\frac{1}{f}\right) \ .
\end{eqnarray}
We consider $d^2 /d\lambda^2 (1/f) =~ $const. in the above equation 
as a simple choice.
This gives
\begin{eqnarray}
f(\lambda) = \frac{1}{1 + \beta \lambda + \gamma \lambda^2} \ ,
\end{eqnarray}
where $\beta , \gamma$ are constant parameters. 
That is, the scale factor is expressed by
\begin{eqnarray}
a(z)=\frac{1}{1+\beta\tanh z+\gamma\tanh^2 z}
\cdot\frac{\ell}{\cosh z} \ .
\label{general-scale}
\end{eqnarray}
Here, we have to impose conditions on parameters 
so that the scale factor is positive.
Substituting this expression into Eq.~(\ref{hc4}), we find the derivative
of the scalar field of this form
\begin{eqnarray}
\kappa^2 \phi^{\prime 2} = \frac{ 4\gamma}{1+\beta\tanh z+\gamma\tanh^2 z}
\cdot\frac{1}{\cosh^4 z}  \ .
\label{quad-phi}
\end{eqnarray}

Thus, we will have three kinds of potentials according to the discriminant
of the quadratic polynomial of $\tanh z$ in the denominator.

\begin{figure}[htbp]
\includegraphics[width=10cm]{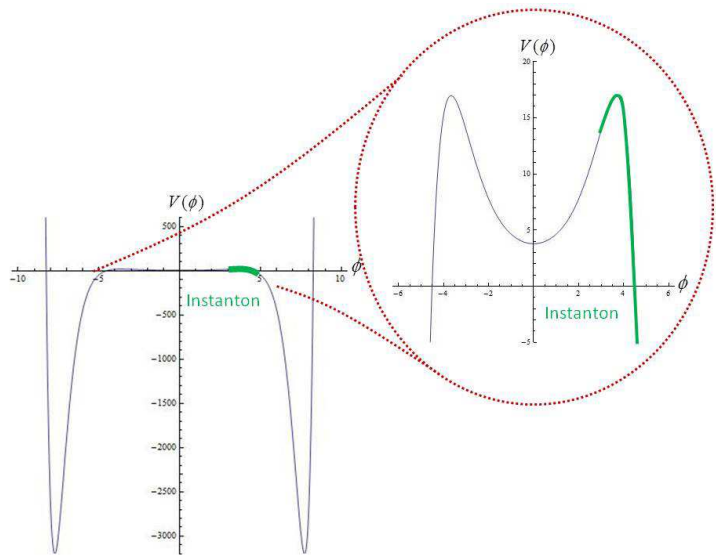}
\caption{The potential, $V$ versus $\phi$ for $\ell=0.25,~\beta=4.1\ell,~\gamma=\ell$.
$\kappa=1$. The region around a metastable dS vacuum is enlarged in a red circle.
}
\label{fig:6}
\end{figure}

\subsubsection{ From dS to AdS }
\label{dsads}

First, we consider the case $\beta^2 -4 \gamma >0$. 
We can write Eq.~(\ref{quad-phi}) as
\begin{eqnarray}
\kappa^2 \phi^{\prime 2} 
= \frac{ 4}{\left(\tanh z + \frac{\beta}{2\gamma} \right)^2
 -\left(\frac{\beta^2}{4\gamma^2} - \frac{1}{\gamma} \right) }
\cdot\frac{1}{\cosh^4 z} \ .
\end{eqnarray}
It is easy to integrate the above equation and obtain the solution
\begin{eqnarray}
\frac{\kappa}{2}\phi=\cosh^{-1}
\left[~
\left(\frac{\beta^2}{4\gamma^2}-\frac{1}{\gamma}\right)^{-\frac{1}{2}}
\left(\tanh z+\frac{\beta}{2\gamma}\right)
~\right] \ .
\label{phi3}
\end{eqnarray}
Since $|\tanh z|<1$, the region of the scalar field is
restricted to $ \frac{2}{\kappa}\cosh^{-1}
\frac{-1+\frac{\beta}{2\gamma}}
{\sqrt{\frac{\beta^2}{4\gamma^2}-\frac{1}{\gamma}}}
< \phi < \frac{2}{\kappa}\cosh^{-1}
\frac{1+\frac{\beta}{2\gamma}}
{\sqrt{\frac{\beta^2}{4\gamma^2}-\frac{1}{\gamma}}}
$.
And the potential is given by
\begin{eqnarray}
\kappa^2\ell^2~V(\phi)
&=&
\frac{1}{128 \gamma^2}\left(\beta^2-4\gamma\right) 
\left[~
115 \beta^2 - 236 \gamma 
- 224 \gamma^2 - 176 \beta 
\sqrt{\beta^2 - 4\gamma}
\cosh\frac{\phi}{2}
\right.\nonumber\\
&&\left.\hspace{1cm}
+ 4 \left(19\beta^2 - 36\gamma - 40 \gamma^2\right)
\cosh\phi 
- 16\beta
\sqrt{\beta^2-4\gamma}
\cosh\frac{3\phi}{2}
+\left(\beta^2- 4\gamma\right)\cosh 2\phi
~\right]   \ ,
\end{eqnarray}
where we used $\tanh z=-\frac{\beta}{2\gamma}
+\sqrt{\frac{\beta^2}{4\gamma^2}-\frac{1}{\gamma}}
\cosh \frac{\kappa}{2}\phi $ in Eq.~(\ref{phi3}) to obtain this result.
The potential is an even function with two AdS local minima and a 
dS metastable 
vacuum as shown in Fig.~\ref{fig:6}.

\subsubsection{ From AdS to AdS}

Next, we consider the case $\beta^2 -4 \gamma <0$. In this case, we can write 
Eq.~(\ref{quad-phi}) as
\begin{eqnarray}
\kappa^2 \phi^{\prime 2} 
= \frac{ 4}{\left(\tanh z + \frac{\beta}{2\gamma} \right)^2
 + \left(  \frac{1}{\gamma} - \frac{\beta^2}{4\gamma^2}  \right) }
\cdot\frac{1}{\cosh^4 z} \ .
\end{eqnarray}
This equation can be integrated as
\begin{eqnarray}
\frac{\kappa}{2}\phi=\sinh^{-1}
\left[~
\left(-\frac{\beta^2}{4\gamma^2}+\frac{1}{\gamma}\right)^{-\frac{1}{2}}
\left(\tanh z+\frac{\beta}{2\gamma}\right)
~\right] \ .
\label{phi4}
\end{eqnarray}
Since $|\tanh z|<1$, the region of the scalar field is
restricted to $ \frac{2}{\kappa}\sinh^{-1}
\frac{-1+\frac{\beta}{2\gamma}}
{\sqrt{\frac{\beta^2}{4\gamma^2}-\frac{1}{\gamma}}}
< \phi < \frac{2}{\kappa}\sinh^{-1}
\frac{1+\frac{\beta}{2\gamma}}
{\sqrt{\frac{\beta^2}{4\gamma^2}-\frac{1}{\gamma}}}
$.
And the potential is given by 
\begin{eqnarray}
\kappa^2\ell^2~V(\phi)
&=&
\frac{1}{128 \gamma^2}\left(\beta^2-4\gamma\right) 
\left[~
115 \beta^2 - 236 \gamma 
- 224 \gamma^2 - 176 \beta 
\sqrt{ 4\gamma -\beta^2}
\sinh \frac{\phi}{2}
\right.\nonumber\\
&&\left.\hspace{1cm}
+ 4 \left( 36\gamma + 40 \gamma^2 -19\beta^2 \right)
\cosh\phi 
+ 16\beta
\sqrt{ 4\gamma -\beta^2}
\sinh\frac{3\phi}{2} 
+\left(\beta^2- 4\gamma\right)\cosh 2\phi
~\right]  \ ,
\end{eqnarray}
where we used $\tanh z=-\frac{\beta}{2\gamma}
+\sqrt{\frac{1}{\gamma} - \frac{\beta^2}{4\gamma^2}}
\sinh\frac{\kappa}{2}\phi$ in Eq.~(\ref{phi4}) in obtaining this result.

\begin{figure}[htbp]
\includegraphics[width=10cm]{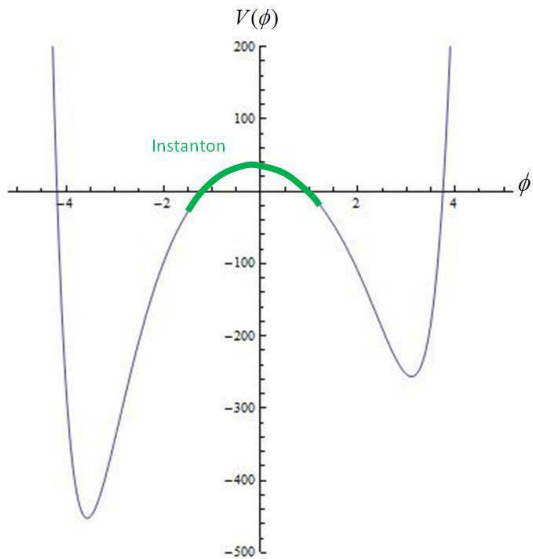}
\caption{The potential $V$ versus $\phi$ for $\ell=1/3,~\beta=-0.5\ell,
~\gamma=1.5\ell$.
$\kappa=1$. 
The green line shows the region of the CDL instanton.}
\label{fig:7}
\end{figure}

This potential has two AdS local minima. We have plotted a typical example 
in Fig.~\ref{fig:7}. The instanton describes vacuum decay from a metastable AdS vacuum to a stable AdS vacuum. The green line shows the range of the instanton.
The endpoints of the green line are in negative energy region with the parameters used in Fig.~\ref{fig:7}. Note that both outside and inside space of the bubble 
when nucleation happened are compact AdS spaces. 
This is possible because the scale factor of the dS space $(=1/\cosh z)$ and 
the AdS space $(=1/\sinh z)$ have the same asymptotic form $(\sim e^{-z})$. 

\subsubsection{runaway type potential}

Finally, we consider the case $\beta^2 -4\gamma = 0$.
In this case, the scale factor becomes
\begin{eqnarray}
a(z)=\frac{4}{\beta^2 \left( \tanh z + \frac{2}{\beta}\right)^2 }
\cdot\frac{\ell}{\cosh z} \ .
\end{eqnarray}
Therefore, this reduces to a special case of the solution in Section 
\ref{type3}.

Note that we find that these instantons do not tunnel from a local minimum 
to a local minimum of the potential same as the runaway type potentials, 
even though the potential has two local minima  now. 
This is the effects of gravity. Thus it is still difficult to read off 
the false and true vacuum of the potential without knowing the whole shape of it. 

After the nucleation of the bubble that we presented above, we have to 
solve Lorentzian equations of motion. The point is that the CDL instantons 
give the initial conditions for this purpose.

\section{Actions of Instantons}

Now that we obtained exact CDL instantons, we can calculate the action of instantons.
Substituting the Hamiltonian constraint equation (\ref{hc1}) 
into the action (\ref{action}), the action in the conformal
coordinate is expressed by
\begin{eqnarray}
S_{E} = 4\pi^2 \int_{M} dz \left[ a^4 V - \frac{3}{\kappa^2}a^2  \right] \ ,
\label{actionwhc}
\end{eqnarray}
where we used $v_{S^3}=2\pi^2$. 
Notice that the surface term is canceled by the GH term. Since all of
our instanton solutions are compact, we do not need $S_b$ in Eq.~(\ref{physical}) 
to make the action well defined. Hence, we have a simple relation $I_{\rm instanton}=S_{E}({\rm instanton})$. 
Substituting an instanton solution  into the action (\ref{actionwhc}), we get 
$I_{\rm instanton}$. Remarkably, we can perform the calculation analytically.

\subsection{From dS to AdS}

Let us start with the case of dS false vacua. 
Putting Eqs.~(\ref{sc3}) and (\ref{general-scale}) into  (\ref{actionwhc}), we can analytically deduce
the action of the instanton solution as
\begin{eqnarray}
I_{\rm instanton} &=& S_E
=- 8\pi^2\frac{\ell^2}{\kappa^2}\left[~
\frac{\beta^2-2\gamma\left(1+\gamma\right)}
{\left(1-\beta+\gamma\right)\left(1+\beta+\gamma\right)
\left( \beta^2 - 4\gamma \right)} \right.\nonumber\\
&& \hskip 2.5cm \left. 
+\frac{\gamma}{\left(\beta^2-4\gamma\right)^{\frac{3}{2}}}
\left\{
\log \left(
\frac{\beta+2\gamma +\sqrt{\beta^2-4\gamma}}
{\beta+2\gamma -\sqrt{\beta^2-4\gamma}}
\right)
- \log \left(
\frac{\beta-2\gamma+\sqrt{\beta^2-4\gamma}}
{\beta-2\gamma-\sqrt{\beta^2-4\gamma}}
\right)
\right\}
~\right] \ .
\end{eqnarray}
Here we assumed the condition for dS false vacua $\beta^2 -4 \gamma >0$. 
In the case of the model with parameters 
$\ell=0.25,~\beta=4.1\ell,~\gamma=\ell$, the value of the action is given by
\begin{eqnarray}
I_{\rm instanton} = -14.0056~\frac{\ell^2}{\kappa^2} \ .
\end{eqnarray}

\subsection{From Runaway to AdS}

In the case of Minkowski false vacua, we have three types of model
as seen in Section \ref{IIIA}. 

In the type $f(\lambda ) = \exp \left( \alpha \lambda \right)$, 
the action of the CDL instanton is analytically expressed by
\begin{eqnarray}
I_{\rm instanton} = S_E= 
-4\pi^2\frac{\ell^2}{\kappa^2}~\frac{\sinh 2\alpha}{\alpha}  \ ,
\end{eqnarray}
where we plugged Eqs.~(\ref{a3}) and (\ref{general-scale}) into (\ref{actionwhc}).

In the case $ f (\lambda) = 1+\beta\lambda$, we get
\begin{eqnarray}
I_{\rm instanton} = S_E
= -8\pi^2\frac{\ell^2}{\kappa^2}\left( 1 + \frac{\beta^2}{3} \right) \ ,
\end{eqnarray}
where we substituted Eqs.~(\ref{a1}) and  (\ref{general-scale}) into (\ref{actionwhc}).

 In more general cases
$f(\lambda) = \left(1 + \gamma \lambda\right)
/\left(\alpha + \beta \lambda\right)^2$, the action is expressed by
\begin{eqnarray}
I_{\rm instanton} = S_E
=-\frac{8\pi^2\ell^2}{3\kappa^2}~\frac{\alpha^2\left(3+\gamma^2\right)
+\beta^2\left(1+3\gamma^2\right)-8\alpha\beta\gamma}{\left(\alpha-\beta\right)^3\left(\alpha+\beta\right)^3}
 \ ,
\end{eqnarray}
where we put Eqs.~(\ref{aa3})  and  (\ref{general-scale}) into  (\ref{actionwhc}).

\subsection{From AdS to AdS}

Similarly, for the AdS false vacua, by using Eqs.~(\ref{general-scale}) and (\ref{sc3}) in  (\ref{actionwhc}), 
we can evaluate the action analytically as 
\begin{eqnarray}
I_{\rm instanton} &=& S_E = 8\pi^2\frac{\ell^2}{\kappa^2}\left[~
\frac{\beta^2-2\gamma\left(1+\gamma\right)}
{\left(1-\beta+\gamma\right)\left(1+\beta+\gamma\right)
\left(4\gamma-\beta^2\right)}  \right. \nonumber\\
&& \hskip 2.5cm \left. 
+\frac{2\gamma}{\left(4\gamma-\beta^2\right)^{\frac{3}{2}}}
\left\{
\tanh^{-1}\left(
\frac{\beta-2\gamma}{\sqrt{4\gamma-\beta^2}}
\right)
-\tanh^{-1}\left(
\frac{\beta+2\gamma}{\sqrt{4\gamma-\beta^2}}
\right)
\right\}
~\right]
\ ,
\end{eqnarray}
where we assumed the condition for AdS false vacua $\beta^2 -4 \gamma <0$.
In the case of the model with parameters $\ell=1/3,~\beta=-0.5\ell,~\gamma=1.5\ell$ in Fig.~\ref{fig:7},
 we obtain 
\begin{eqnarray}
I_{\rm instanton}= -6.83904~\frac{\ell^2}{\kappa^2} \ .
\end{eqnarray}

\section{Exact Decay Rates of False Vacuum with Gravity }
\label{subtlety}

As we obtained the exact CDL instantons and their actions, we are now ready to calculate decay rates. 
According to the prescription by Coleman and De Luccia, the decay rate
can be evaluated by the formula
\begin{eqnarray}
\Gamma = A e^{-B} \ ,
\end{eqnarray}
where the prefactor $A$ is derived from quantum corrections and
$B$ consists of the difference between the Euclidean action for the CDL instanton solution, $S_E({\rm instanton})$,
and that of the false vacuum, $S_f$. In defining $B$, we also adopt the GH prescription (\ref{physical}) to make the action well defined.
Thus, the bounce action is expressed by $B= I_{\rm instanton} -I_{f}$ where
$I$ is defined by Eq.~(\ref{physical}) for non-compact geometries and hence $I$ simply becomes 
$S_E$ for compact ones. 
In the original work by Coleman and De Luccia, the thin-wall approximation 
has been used.
In this approximation, instantons connect two vacuum states. Hence, 
the terms of $S_b$ coming from each $I_{\rm instanton}$ and $I_f$ are canceled out
even for the non-compact false vacua. Then
the decay rate is always calculated simply by taking the difference of the action 
for the CDL instanton, $S_E({\rm instanton})$, 
and that of the false vacuum $S_f$. 
The above definition of $B$ is equivalent to this original CDL definition
 in the thin-wall approximation. 
In the presence of gravity, however, we cannot use the thin-wall approximation. And,
the asymptotic value of the scalar field of the CDL instanton does not necessarily
coincide with that in the false vacuum. This fact makes
the difference in evaluation of decay rate when 
the false vacuum is non-compact geometry such as Minkowski space and AdS space
as we will see below. 

\subsection{From dS to AdS -- Compact False Vacua}

If the false vacua are compact geometries, we get a conventional result of
decay rate as below.
Let us start with the case that a false vacuum has a positive energy, that is,
a dS vacuum. 
In the case of the four-dimensional Euclidean dS spaces,
 the Hamiltonian constraint equation gives the solution expressed by
\begin{eqnarray}
a = \frac{\ell_{\rm dS}}{\cosh z} \ , \quad
{\rm where} \quad \ell_{\rm dS} = \sqrt{\frac{3}{\kappa^2 V}} \ .
\end{eqnarray}
Here $V(\phi)$ is a positive constant and we took $d\phi/dz=0$ in the
Hamiltonian constraint equation.
Substituting this false vacuum solution into the action Eq.~(\ref{actionwhc}), 
we have
\begin{eqnarray}
I_f = S_f &=& 4\pi^2  \int^{\infty}_{-\infty} dz 
\left[a^4 V - \frac{3}{\kappa^2} a^2 \right] 
= \frac{12\pi^2 \ell_{\rm dS}^2}{\kappa^2} \int^{\infty}_{-\infty} dz 
          \left[ \frac{1}{\cosh^4 z} - \frac{1}{\cosh^2 z}\right] \nonumber\\
    &=& - \frac{24\pi^2}{\kappa^4V}  \ .
\label{sb:ds}
\end{eqnarray}
Note that because dS space is compact, we do not need $S_b$ in Eq.~(\ref{physical}) 
to make the action well defined and then $I_f$ becomes $S_f$.
Using our solution obtained in Section \ref{dsads}, 
we can read off $\kappa^2 \ell^2 V(\phi)=3.8$ from Fig.~\ref{fig:6} in the case of 
$\ell=0.25,~\beta=4.1\ell,~\gamma=\ell$. Then we can evaluate 
the action of the false vacuum as 
\begin{eqnarray}
I_f = -\frac{24\pi^2}{\kappa^4V(\phi)}= -62.3343~ 
\frac{\ell^2}{\kappa^2}\ .
\end{eqnarray}
Thus, we obtain
\begin{eqnarray}
&& B = I_{\rm instanton} - I_f = 48.3287~\frac{\ell^2}{\kappa^2} \ .
\end{eqnarray}
Finally, the decay rate is given by
\begin{eqnarray}
\Gamma = Ae^{-48.3287~\frac{\ell^2}{\kappa^2}} \ .
\end{eqnarray}
We get the decay rate with an exponential suppression in this case
as expected. Remarkably, we could perform an exact calculation
of the tunneling rate with gravity for the first time. 

\subsection{From Runaway to AdS and AdS to AdS -- Non-compact False Vacua}

If the false vacua have non-compact geometries, there exist some subtleties of
the interpretation of decay rates as follows. 
Let us first see the decay of a runaway state into an AdS vacuum.
In this case, the false vacuum can be regarded as a Minkowski vacuum. 
In the Minkowski background, we have 
$V(\phi)=0,~d\phi/dz =0,~ a(z)=\ell_M e^z$,
where $\ell_M $ has the dimension of length.
Then naively, the action itself diverges because of the non-compact geometry 
such as
\begin{eqnarray}
S_f = 4\pi^2 \int^{\infty}_{-\infty} dz
\left[~a^4V(\phi)-\frac{3}{\kappa^2}a^2~\right]=-\infty 
\ .
\label{sb:minkowski}
\end{eqnarray}
However, according to the GH prescription (\ref{physical}), 
this divergence has to be canceled by $K_0$ or equivalently $S_b$. 
Hence, we obtain
\begin{eqnarray}
I_f = S_f - S_b = 0\,,
\end{eqnarray}
and the decay rate is determined only by the instanton action. 

Adopting tentatively the above interpretation,
let us calculate decay rates for the exact solutions 
we obtained in Section \ref{IIIA}.
For the model $f(\lambda ) = \exp \left( \alpha \lambda \right)$, the decay rate becomes
\begin{eqnarray}
    \Gamma = A \exp\left(4\pi^2\frac{\ell^2}{\kappa^2}~\frac{\sinh 2\alpha}{\alpha} \right) \ .
\end{eqnarray}
With the parameter $\alpha=0.1$ in Fig.~\ref{fig:3}, we have
\begin{eqnarray}
\Gamma = A e^{79.4843~\frac{\ell^2}{\kappa^2}}  \ .
\end{eqnarray}
In the case $ f (\lambda) = 1+\beta\lambda$, the decay rate becomes
\begin{eqnarray}
  \Gamma = A \exp\left[~8\pi^2 \frac{\ell^2}{\kappa^2}
\left( 1 + \frac{\beta^2}{3} \right)~\right] \ .
\end{eqnarray}
With the parameters $\ell=3.8,~\ell\beta=0.8$ in Fig.~\ref{fig:4}, we have 
\begin{eqnarray}
\Gamma = A e^{1156.98~\frac{\ell^2}{\kappa^2}}  \ .
\end{eqnarray}
Finally, the model
$f(\lambda) = \left(1 + \gamma \lambda\right)
/\left(\alpha + \beta \lambda\right)^2$
with the parameters $\alpha=3,~\beta=1,~\ell=2,~\ell\gamma=1$ in Fig.~\ref{fig:5}, 
the decay rate reads 
\begin{eqnarray}
\Gamma=A e^{3.90672~\frac{\ell^2}{\kappa^2}}\,.
\end{eqnarray}
Apparently, these results are unconventional. 
Indeed, the decay rate is not exponentially suppressed at all. 

However, the result will completely change if we regard
flat space as the infinite radius limit of dS space. 
That is, we use the dS action (\ref{sb:ds})
instead of the action (\ref{sb:minkowski}). 
In the limit of $V=0$, $I_f$ becomes minus infinity. This makes 
the decay rate vanish. Thus, the continuity argument
 leads to the result that the decay rate through the CDL instanton 
is zero~\cite{Aguirre:2006ap,Bousso:2006am}. 
We can also argue that introducing $K_0$ or equivalently $S_b$
 may not be appropriate
 in our situation.  This argument leads to the same conclusion that the
decay rate becomes zero, and is consistent with the above continuity 
argument.

In the AdS background, $V(\phi)<0,~d\phi/dz =0,~ a(z)=\ell_{\rm AdS}/\sinh z$.
The Hamiltonian constraint equation 
gives $\ell_{\rm AdS}=\sqrt{-3/\kappa^2V(\phi)}$. 
Then the value of the action is given by
\begin{eqnarray}
S_f = 4\pi^2 \int^{\infty}_{-\infty} dz
\left[~a^4V(\phi)-\frac{3}{\kappa^2}a^2~\right]=-\infty\,,
\qquad
I_f = S_f - S_b = 0\,,
\label{sb:ads}
\end{eqnarray}
where the divergence is assumed to be canceled by the 
GH prescription (\ref{physical}).
This corresponds to the unconventional interpretation
discussed above, and again the decay rate would not be suppressed.

On the other hand, as we have argued previously,
 we may discard $S_b$ from the action for AdS space in Eq.~(\ref{physical}).
 Under this prescription, the action for AdS vacuum diverges negatively
 and the decay rate vanishes. 
This is the conventional interpretation.

At the moment, however, we feel 
it is still premature to extract definite conclusion from 
the thick-wall CDL instantons.

\section{Conclusion}

Motivated by the string theory landscape, we have studied the vacuum decay into
AdS vacuum with gravity. 
We have succeeded in obtaining exactly solvable CDL instantons.
We focused on the fact that the CDL instantons can be obtained 
by deforming HM instantons. We first gave a scale factor of a deformed
HM instanton.
Once a scale factor is given, the scalar field and the 
potential function  can be obtained as a function of the 
radial coordinate. Combining them, we analytically obtained the potential 
function for the deformed HM instanton. 
In this way, we have obtained exact CDL instantons systematically.
As a result, we have constructed exactly solvable models corresponding to
 the decay process from dS, AdS and Minkowski spaces into AdS spaces. 
It is known that there exist either compact or non-compact CDL instantons
depending on the shape 
of the potential~\cite{Aguirre:2006ap,Bousso:2006am}.
In this paper, as we focused only on compact instantons, we did not 
have to be bothered with non-compact CDL instantons.
In principle, however, we can find a variety of exact solutions including non-compact CDL instantons
by extending our method~\cite{Kanno:2012zf}. 

As we got exact CDL instantons, we also succeeded in evaluating the action  
for the decay rate analytically.
We have calculated the decay rate for the vacuum decay into the AdS vacuum.
We have also revealed a subtle point in evaluating decay rates for thick-wall
CDL instantons. 

All of the decay processes we have considered go to an AdS crunch 
singularity~\cite{Coleman:1980aw}.
This would be a problem in the string theory landscape. A proposal of 
holographic excision~\cite{Maldacena:2010un} may help to cure this singularity.
It is interesting to investigate the excision~\cite{Harlow:2010my,Garriga:2010fu,Kanno:2011hs} in our exactly solvable models.

\acknowledgments
We would like to thank Jose Blanco-Pillado, Larry Ford, Hideo Kodama, 
Ben Shlaer, and especially Misao Sasaki and Alex Vilenkin for useful 
and stimulating discussions.
SK was supported in part 
by grant PHY-0855447 from the National Science 
Foundation.
JS was supported in part by the
Grant-in-Aid for  Scientific Research Fund of the Ministry of 
Education, Science and Culture of Japan No.22540274, the Grant-in-Aid
for Scientific Research (A) (No.21244033, No.22244030), the
Grant-in-Aid for  Scientific Research on Innovative Area No.21111006,
JSPS under the Japan-Russia Research Cooperative Program,
the Grant-in-Aid for the Global COE Program 
``The Next Generation of Physics, Spun from Universality and Emergence".

\end{document}